

\baselineskip=20pt
\magnification=1200
\hsize 15.5truecm\hoffset 1.2truecm
\vsize 22.0truecm\voffset 1.5truecm
\outer\def\beginsection#1\par{\medbreak\bigskip
      \message{#1}\centerline{\bf#1}\nobreak\medskip\vskip-\parskip
      \noindent}
\font\grande=cmbx10 scaled\magstep1

\def \La {\Lambda}

\def \a {\alpha}

\def \da {\delta}

\def \th {\theta}
\def \Sg {\Sigma}
\def \vphi {\varphi}

\def \rline {\rightline}
\def \lline {\leftline}
\def\sitontop#1#2{\mathrel{\mathop{#1}\limits^{\scriptscriptstyle #2}}}
\def\tr{{\rm Tr}}
\def\um{{1\over2}}
\def\o#1#2{{#1\over#2}}

\def\pa{\partial}

\rline {DFTT 21/95}
\rline {gr-qc/9504001}
\rline {March 1995}
\centerline{\grande $\La \to 0$ limit of 2+1 Quantum Gravity for
arbitrary genus }
\vskip 0.5truecm
\centerline {J.E.Nelson}
\centerline {Dipartimento di Fisica Teorica dell'Universit\`a degli
Studi di Torino}
\centerline {Via Pietro Giuria 1, 10125,Torino, Italy}
\centerline{\it email:~ nelson@to.infn.it,~ telefax ~039-11-6707214}
\vskip 0.5truecm
\centerline{T.Regge}
\centerline {Dipartimento di Fisica del Politecnico di Torino}
\centerline {Corso Duca degli Abruzzi 24, 10129,Torino, Italy}
\centerline{\it email:~ regge@to.infn.it,~ telefax ~039-11-5647399}

\beginsection Abstract

The abstract quantum algebra of observables for 2+1 gravity is
analysed in the limit of small cosmological constant. The algebra splits
into two sets with an explicit phase space representation;~one set
consists of $6g-6$ {\it commuting} elements which form a
basis for an algebraic manifold defined by the trace and rank
identities;~the other set consists of $6g-6$ tangent vectors to this
manifold. The action of the quantum mapping class group leaves the
algebra and algebraic manifold invariant. The previously presented
representation for $g=2$ is analysed in this limit and reduced to
a very simple form. The symplectic form for $g=2$ is computed.
\lline{P.A.C.S. 04.60}
\vfill

\beginsection 1. Introduction.

For some time we have been analysing the abstract quantum commutator
algebra for $2+1$ gravity with negative cosmological constant $\La$ [1,2]:
$$
\eqalignno{
[a_{mk},a_{jl}]\ &=\ [a_{mj},a_{kl}]\ =\ 0 & (1.1) \cr
[a_{jk},a_{km}]\ &=\ \Big(\o {1} {K}-1\Big)(a_{jm}-a_{jk}a_{km}) & (1.2)
\cr
[a_{jk},a_{kl}]\ &=\ \Big(1-\o {1} {K}\Big)(a_{jl}-a_{kl}a_{jk}) & (1.3)
\cr
[a_{jk},a_{lm}]\ &=\ \Big(K-\o {1} {K}\Big)(a_{jl}a_{km}-a_{kl}a_{jm}) &
 (1.4) \cr}
$$
where $K\ =~ e^{-2i\th}, \tan \th =-{\hbar \over 8\a}, \La\ =\ -\o {1}
{\a^{2}}$ is the cosmological constant and $\hbar$ is Planck's constant.
In (1.1-4) $m,j,l,k$. are 4 anticlockwise points of a $n-gon$ (see [1])
labelled anticlockwise $m,j,l,k\ =\
1\cdots n$, and the time independent quantum operators $a_{lk}$
correspond to the classical $\o {n(n-1)} {2}$ gauge invariant trace
elements
$$
\a_{ij}\ =\ \a_{ji}\ =\ \um\tr\Big(S(t_{i}t_{i+1}\cdots t_{j-1})\Big)\ ,
\ S\in SL(2,R)
\eqno (1.5)
$$
The map $S\ :\ \pi_{1}(\Sg)\sitontop {\longrightarrow} {S} SL(2,R)$ is
defined by the integrated anti-De~Sitter connection in the initial data
Riemann surface $\Sg$ of genus $g$ when $n\ =\ 2g+2$, and refers to one
of the two spinor components, say the upper component, of the spinor
group $SL(2,R)\otimes SL(2,R)$ of the gauge group $SO(2,2)$ of $2+1$
gravity with negative cosmological constant [2]. The lower component
yields an independent algebra of traces $b_{ij}$ identical to (1.1-4)
but with $K\to 1/K$. Moreover $[a_{ij},b_{kl}]\ =\ 0\ \forall\ i,j,k,l$.

The algebra (1.1-4) is invariant under the quantum action of the mapping
class group on traces [1], provided that the operators
in (1.1-4) are ordered with the convention that $d(a_{ij})$ is
increasing from left to right where $d(a_{ij})\ =\ \o {(i-1)(2n-2-i)} {2}
+j-1$.

In previous articles we only
considered the upper component, which we now summarise. The homotopy
group $\pi_{1}(\Sg)$ of the surface is defined [1]
by generators $t_{i},\ i\ =\ 1\cdots 2g+2$ and presentation:
$$
t_{1}t_{3}\cdots t_{2g+1}\ =\ 1\ ,\ t_{2}t_{4}\cdots t_{2g+2}\ =\ 1 ,
t_{1}t_{2}\cdots t_{2g+2}\ =\ 1\ \eqno (1.6)
$$
The last relator in (1.6) implies that $\Sg$ is closed.

The algebra (1.1-4) seems overprescribed since the number of elements
$a_{ij}$ is $\o {n(n-1)} {2} = (g+1)(2g+1)$. In [3] we determined for
$n\leq 6$, i.e. $g\leq 2$
a set of $p$ linearly independent central elements $A_{nm},\ m\ =\ 1
\cdots p$ where $n\ =\ 2p$ or $n\ =\ 2p+1$, and analysed the trace
identities which follow from the presentation (1.6) of the homotopy
group $\pi_{1}(\Sg)$ and a set of rank identities. These identities
together generate a two-sided ideal, from which it was deduced that for
generic $g$ there are
precisely $6g-6$ independent elements for each $(\pm)$ algebra.
The reduction from $\o {n(n-1)} {2}\ =\ (g+1)(2g+1)$ to $6g-6$ results
from the use of the above mentioned identities  but is highly non
unique. In [4] this was implemented explicitly for $g=2$ reducing the
number of independent variables for the (+) algebra from 15 to 6, and
the representations discussed.

In this paper we analyse this quantum theory (the traces $a_{ij}$
and the commutator algebra (1.1-4)) in the limit of small
cosmological constant through an expansion around $\sqrt -\La =0$,
for arbitrary genus $g$.
Note that the theory was first formulated for $\La=0$ [5], and then
generalised. In [6] similar results were obtained for $g=1$ by first
taking the limit of the classical (Poisson bracket) algebra and then
quantising. Only traces corresponding to paths with intersections
$0,\pm 1$ (eq.1.1-3) were considered. The limit as $\La \to 0$ of the
exact classical solution for $g=1$ was also studied in [7]. Here we show
that for arbitrary genus $g$ the two sets (the $a_{ij}$ and $b_{ij}$ or
the $\pm$) of $6g-6~$ variables split into another two $6g-6$ sets
with an explicit phase space representation. One of these consists
of $6g-6$ {\it commuting} elements which form a
basis for an algebraic manifold defined by the trace and rank
identities, the other set consists of $6g-6$ tangent vectors. These
and the action of the quantum mapping class group which leaves the
algebra and algebraic manifold invariant are presented in Section 2.
A representation is presented in Section 3. In Section 4 the previously
presented
representation for $g=2$ is analysed in this limit and reduced to
a very simple form, and the symplectic form for $g=2$ is computed.

\beginsection 2. Separation of variables for small $\La$

To discuss the limit of small cosmological constant it is necessary, as
for g=1 [6,7], to consider simultaneously the upper and lower spinor
components. From now on denote $a_{ik}$  by ${a^+}_{ik}$ and $b_{ik}$ by
${a^-}_{ik}$
and set
$$
{a^{\pm}}_{ij}(\th) =s_{ij} \pm \th~ t_{ij}  +O(\th^2) \eqno (2.1)
$$
with $s_{ij} ~=~s_{ji},~t_{ij}~=~t_{ji}$ independent of $\th$ and
$s_{ii} =1, t_{ii}=0$. The expression (2.1) therefore corresponds to
an expansion of the
$a_{ij}$ to $O(\La)$ around $\La=0,~K=1,~\th=0$. The algebra of these
variables can be calculated directly from (1.1-4) or by noting that
they can be expressed as
$$
{a^{\pm}}_{ij}(0)= s_{ij}, \quad
{{d {a^{\pm}}_{ij}} \over {d \th}}(0)= ~\pm~ t_{ij} \eqno (2.2)
$$
By repeated differentiation with respect to $\th$ in the limit $\th
\to 0$
it follows from (1.1-4) that all the $s_{ij}$ commute amongst
themselves whereas the $s$ and $t$ variables satisfy the commutators
$$
\eqalignno{
[s_{jk},t_{km}]\ &=\ -[s_{km},t_{jk}] = i~(s_{mj}- s_{jk}s_{km}) & (2.3) \cr
[t_{mj},t_{jl}]\ &= i~(t_{ml} - t_{jl}s_{mj} - \ t_{mj}s_{jl} ) \cr
               \ &= i~(t_{ml} - s_{mj}t_{jl} - \ s_{jl}t_{mj})  & (2.4) \cr
[s_{ml},t_{jk}]\ &=\ -[s_{jk},t_{ml}] = 2i~(s_{jl}s_{mk} - s_{mj}s_{kl}) &
(2.5) \cr
[t_{ml},t_{jk}]\ &=\ 2i~( s_{jl}t_{mk} - s_{mj}t_{kl}
-  s_{kl}t_{mj} + \ s_{mk}t_{jl}) & (2.6) \cr}
$$
with all other commutators zero. As for the algebra (1.1-4), m,j,l,k are
any 4 anticlockwise points (e.g.1,2,3,4 or $2g+2$,1,4,5 etc) of the
polygon representation (see [1]). Some comments are in order.

The first two commutators (2.3-4) correspond
to paths on $\Sg$ with single intersections, already reported in [5]
\footnote {$^{*}$}{The commutators (2.3-4) were first computed for $g=1$
in [5]. Equations (2.3-4) here correspond to equations (5.4-5) of [5]
with the following identifications; the variables $q,\nu$ of [5]
correspond here to
$s,t$ respectively and the labelling of the paths $jk,km$ (or $mj,jl$)
on $\Sg$ to $u,v$. To get precisely equations (2.3-4) it is necessary to
use the $SL(2,R)$ identities
$$\nu(uv^{-1}) +\nu(uv)=2 (q(v) \nu(u) + q(u) \nu(v)),  \qquad
q(uv^{-1}) + q(uv)= 2 q(u) q(v).$$ }.
The last two commutators (2.5-6) correspond to paths on $\Sg$ with
double intersections, which could be determined from the Jacobi
identity which is satisfied for all triples. The
first equalities of (2.3) and (2.5) follow from
$$
[{a^+}_{jk},{a^-}_{km}] = [{a^+}_{ml},{a^-}_{jk}] = 0
$$
whereas the commutators (2.4) and (2.6) between two $t$'s can be
deduced from the Jacobi identity on a $(t,t,s)$ triple.

The commuting $s$ variables satisfy the same trace and rank
identities as the $a_{ij}$ variables [3] with $\th=0$. Given just one
identity all others can be obtained
by repeated commutation with the elements $t_{ij}$ of the algebra
(2.3-6), or equivalently, by repeated application of
$$
[t_{uv},s_{jk}]~{\pa I(s) \over {\pa s_{uv}}}
$$
(to be summed over $u,v$). The set of commuting $s$ variables can be
used as a basis for the algebraic manifold defined by $I(s)$ ~=~0.

Similar identities for the $t$ variables are obtained from the $s$
identities $I(s)$ as
$$
I(t) = {\pa I(s) \over {\pa s_{uv}}} t_{uv}
$$
(again, to be summed over all $u,v$). It
follows that the number of $t$ identities $I(t)$ is equal to the number
of $s$ identities $I(s)$ so that there are equal numbers of independent
$s$ and $t$ variables.
These identities are certainly not all independent. In fact they form an
ideal under commutation.

Under the action of the quantum Dehn group with elements
$D_{ij}=D_{ji}, i,j=1 \cdots 2g+2 $ the $s$ and $t$ variables transform
as follows:
$$
\eqalignno{
D_{ml}:t_{ml}  \to {t \prime}_{ml}~=~ & t_{ml} \cr
       t_{kl}  \to {t \prime}_{kl}~=~ & t_{mk} \cr
       t_{mj}  \to {t \prime}_{mj}~=~ & t_{jl} \cr
       t_{mk}  \to {t \prime}_{mk}~=~ & 2(s_{mk}t_{ml} + s_{ml}t_{mk})
                                                  - t_{lk} \cr
       t_{jl}  \to {t \prime}_{jl}~=~ & 2(s_{jl}t_{ml} + s_{ml}t_{jl})
                                               - t_{mj} \cr
       t_{jk}  \to {t \prime}_{jk}~=~ & -2(s_{mj}t_{mk} + s_{mk}t_{mj}
                                 + s_{jl}t_{kl} + s_{kl}t_{jl})  \cr
& +4(s_{ml}s_{mk}t_{jl} + s_{ml}s_{jl}t_{mk} + s_{jl}s_{mk}t_{ml})
                                                  + t_{jk} \cr
       s_{ml}  \to {s \prime}_{ml}~=~ & s_{ml} \cr
       s_{kl}  \to {s \prime}_{kl}~=~ & s_{mk} \cr
       s_{mj}  \to {s \prime}_{mj}~=~ & s_{jl} \cr
       s_{mk}  \to {s \prime}_{mk}~=~ & 2s_{ml}s_{mk} - s_{kl} \cr
       s_{jl}  \to {s \prime}_{jl}~=~ & 2s_{ml}s_{jl}  - s_{mj} \cr
       s_{jk}  \to {s \prime}_{jk}~=~ & - 2(s_{mj}s_{mk} + s_{kl}s_{jl})
                           + 4s_{ml}s_{jl}s_{mk} + s_{jk} & (2.7) \cr } $$
and leave the algebra (2.3-6)  and the ideal of $s$ and $t$ identities
invariant. The transformations (2.7) can be simplified by noting that
$$
{t \prime}_{pq}~=~ {{\pa {s \prime}_{pq}} \over {\pa s_{uv}}} t_{uv}
\qquad (sum~ over~ u,v)
$$
for generic indices (points of the polygon) $p,q,u,v$. The $D_{ml}$
as expressed in (2.7) satisfy the identities of the Braid group
$B(2g+2)$ [8,9].
\beginsection 3. Representations.

The algebra (2.3-6) admits the following representation;

Let the set of the basis $s_{ij}$ variables act as configuration space
variables by
multiplication, and let the $t_{ij}$ variables act by differentiation
$$
t_{ij} = C_{ij,kl}(s) {\pa \over {\pa s_{kl}}} \qquad (sum~ over~ k,l)
\eqno(3.1)$$
where
$$
C_{ij,kl}(s)~=~[t_{ij},s_{kl}]~=~-~C_{kl,ij}(s)
$$
is at most quadratic in $s$. Therefore the $t_{ij}$ can be considered
as tangent vectors to the algebraic manifold with basis $s_{ij}$.

\beginsection 4. Representation for g = 2.

For $g = 2$, each $(\pm)$ algebra (1.1-4) consists of 15 elements
$a^{\pm}_{ij}$ but by
use of computer algebra we were able [4] to reduce to $6~=~6g-6$
independent variables for each $(\pm)$ as follows, by satisfying all
the trace and rank identities. A convenient choice for the 6
independent elements is given by 6 commuting angles
$\vphi^{\pm}_{-b}\ =\ -\vphi^{\pm}_{b},\ b\ =\ \pm 1\cdots\pm 3$
with
$$\vphi^+_a(K) =\vphi^-_a(\o {1} {K})$$
and
$$
a^{\pm}_{12}\ =\ \o {\cos\vphi^{\pm}_{1}} {\cos\th},\qquad
a^{\pm}_{34}\ =\ \o {\cos\vphi^{\pm}_{2}} {\cos\th}  \qquad
a^{\pm}_{56}\ =\ \o {\cos\vphi^{\pm}_{3}} {\cos\th}  \eqno (4.1)
$$
with corresponding commuting momenta $\pi^{\pm}_{a}, a=1,2,3$
with
$$\pi^+_a(K) =\pi^-_a(\o {1} {K})$$
and the only non-zero commutators
$$
[\vphi^{\pm}_{a},\pi^{\pm}_{b}]\ =~\pm 2 i \th \da_{ab},\ a,b,\ =\ 1,2,3
\eqno (4.2)
$$
In the following the single indices 1,2,3 refer to the three commuting
sectors 12, 34, 56 whereas the double indices refer to two points of the
hexagon.

It can be checked that the 24 remaining $a^{\pm}_{ik}$ can be expressed
in terms of the $\vphi^{\pm}_{a}$ and their conjugate momenta
$\pi^{\pm}_{a}$. For example the quantum, ordered operator $a^+_{23}$
can be expressed as
$$
\eqalignno{
a^+_{23}\ =\ &\cos \pi^+_1 \cos \pi^+_2 \cr
&\ +\big(\cot \vphi^+_1 \cot \vphi^+_2 -{\cos \vphi^+_3 \over {\cos
\th  \sin \vphi^+_1 \sin \vphi^+_2}} \big) \sin \pi^+_1 \sin \pi^+_2\cr
&\ - i\tan \th \left(\cot \vphi^+_2 \cos \pi^+_1\sin \pi^+_2 +
\cot \vphi^+_1 \cos \pi^+_2\sin \pi^+_1 \right) & (4.3) \cr}
$$
so that {\it all} the 30 $a^{\pm}_{ik}$ are functions of $K\
=~ e^{-2i\th}$, and the six conjugate pairs $\vphi^{\pm}_a,
\pi^{\pm}_a, a =1,2,3$. In [4] it was shown that the requirement that
this representation of all the $a^{\pm}_{ik}$ by hermitian operators
determines the norm in the Hilbert space spanned by the $\cos \vphi_a$,
and restricts the range of the $\vphi_a$. There is some evidence [10-11]
that thr trace holonomies (4.1) should be unbounded (hyperbolic) in
agreement with the partial results of [4].

In the limit $\La \to 0$ the relationship of these variables to those of
Section 1 is as follows.
Let
$$\vphi^{\pm}_a = Q_a \mp \th~ p_a, \qquad \pi^{\pm}_b = q_b \pm \th~ P_b
\eqno(4.4)$$
where all $Q_a, q_a, P_b$ and $p_b$ are independent of $\th$
\footnote {$^{*}$}
{A similar phenomenon occurs for $g=1$ [7]. Here the combinations
$Q_a=\o{\vphi^+_a +\vphi^-_a}{2},~p_a=\o{\vphi^-_a -\vphi^+_a}{2 \th},
{}~q_a=\o{\pi^+_a +\pi^-_a}{2}$ and $P_a=\o{\pi^+_a -\pi^-_a}{2 \th}$
remain finite and constant in the limit $\th \to 0$.}
, given by
$$Q_a = \vphi^+_a(0) = \vphi^-_a(0), \qquad p_a = - \o {\pa
\vphi^+_a}{\pa \th}(0) = \o {\pa \vphi^-_a}{\pa \th}(0)\eqno(4.5)
$$
Their conjugate variables are, respectively,
$$P_b = \o {\pa \pi^+_b}{\pa \th}(0)= -\o {\pa \pi^-_b}{\pa \th}(0)
  \qquad  q_b = \pi^+_b(0) =\pi^-_b(0) \eqno(4.6)
$$
with the only non-zero commutators following from (4.2)
$$
[P_a, Q_b] =[p_b,q_a] = -i \da_{ab}\eqno(4.7)
$$
so that from (4.1) we
have directly the six independent variables\footnote {$^{**}$}
{Three of the six independent variables of [12] for $\La = 0$ and $g=2$
should be compared with (4.8). These correspond, in our notation, to the
traces of the representation of elements 16, 25 and 34 of the hexagon
(see [4] for details). They can be obtained from our 12, 34, and 56
variables by the Dehn transformation $D_{15}$ (eq.(2.7)).}
$$
s_{12}\ =\ \cos Q_1,\qquad
s_{34}\ =\ \cos Q_2\qquad
s_{56}\ =\ \cos Q_3   \eqno (4.8)
$$
and
$$
t_{12}\ =  p_1 \sin Q_1, \qquad
t_{34}\ =  p_2 \sin Q_2, \qquad
t_{56}\ =  p_3 \sin Q_3   \eqno (4.9)
$$
All of the commuting $s_{ij}$ variables are expressed in terms of the
$q_a$ and $Q_a$, the notation has obviously been chosen because of its
suggestive nature e.g. $s_{23}$ is, from (4.3) given by
$$
s_{23} = \cos q_1 \cos q_2 + T_{12,3} \sin q_1 \sin q_2  \eqno(4.10)
$$
where
$$
T_{ij,k}= \cot Q_i \cot Q_j -{\cos Q_k \over {\sin Q_i \sin Q_j}}
\qquad i,j,k~cyclical.
$$
Note that {\it all} of the  $s_{ij}$  can be expressed in this form,
e.g. from (4.8)
$$
s_{12} = \cos Q_1 = \cos Q_2 \cos Q_3 - T_{23,1} \sin Q_2 \sin Q_3
$$
and
$$
s_{13} = \cos (q_1 + Q_1) \cos q_2 + T_{12,3} \sin (q_1 + Q_1)\sin q_2
$$
$$
s_{24} = \cos q_1 \cos (q_2 - Q_2) + T_{12,3} \sin q_1 \sin (q_2 - Q_2)
$$
$$
s_{14} = \cos (q_1 + Q_1) \cos (q_2 - Q_2) + T_{12,3} \sin (q_1 + Q_1)
\sin (q_2 - Q_2)
$$
$$
s_{25} = \cos (q_1 - Q_1) \cos (q_3 + Q_3) + T_{31,2} \sin (q_1 - Q_1)
\sin (q_3 + Q_3)
$$
$$
s_{36} = \cos (q_2 + Q_2) \cos (q_3 - Q_3) + T_{23,1} \sin (q_2 + Q_2)
\sin (q_3 - Q_3)
$$
These variables have striking familiarity and are clearly related to
those presented in the Appendix of [13].
All of the remaining $s_{ij}$ can be obtained from the above by
cyclical rotation of the sector indices 1,2,3, which corresponds to a
cyclical rotation of the hexagon by 2 points (e.g. $s_{35}$ is obtained
from $s_{13}$ etc).

A general formula for the operators $t_{ij}$ in terms of the
operators $Q_a, q_a, P_a, p_a$, ordered with all momenta to the right
and consistent with (3.1) is
$$
t_{ij}= i~\big( \o {\pa s_{ij}}{\pa Q_a} \o{\pa}{\pa q_a}
 - \o {\pa s_{ij}}{\pa q_a} \o{\pa}{\pa Q_a} \big)
= \o {\pa s_{ij}}{\pa q_a} P_a
- \o {\pa s_{ij}}{\pa Q_a}p_a  \eqno(4.11)
$$
when  the commutators (4.7) are represented by
$$
p_a =-i \o {\pa}{\pa q_a}, \qquad P_a= -i \o {\pa}{\pa Q_a}\eqno(4.12)
$$
One example is
$$
\eqalign{&t_{23} = \cr
&\big (\o {T_{12,3}\cos q_1 \sin q_2}{\sin Q_1 \sin Q_2}-
\cos q_2 \sin q_1 \big ) P_1 + \big (\o {T_{12,3}\cos q_2 \sin q_1}
{\sin Q_1 \sin Q_2}- \cos q_1 \sin q_2 \big ) P_2 \cr
&- \o {T_{13,2}\sin q_1 \sin q_2}{{\sin}^2 Q_1 \sin Q_2} p_1
- \o {T_{23,1}\sin q_1 \sin q_2}{\sin Q_1 {\sin}^2 Q_2} p_2
- \o {\sin q_1 \sin q_2 \sin Q_3}{\sin Q_1 \sin Q_2} p_3 \cr}
$$

Clearly (4.11) gives (4.9) for $ij=12,34,56$, and
$$
[t_{ij},s_{ij}]= - i \big(\o {\pa s_{ij}}{\pa Q_a} \o{\pa s_{ij}}{\pa q_a}
- \o {\pa s_{ij}}{\pa q_a} \o{\pa s_{ij}}{\pa Q_a}\big) =0
$$

It can be checked that the algebra (2.3-6) and the identities $I(s)$
and $I(t)$ are satisfied by all 15 $s_{ij}$ and all 15 $t_{ij}$.
The identities $I(s)$ can all be derived from, for example
$$
\eqalign{&s_{12}s_{34}+s_{23}s_{14}-s_{13}s_{24}- s_{56}\ =\ 0 \cr
&s_{12}s_{46}+s_{24}s_{16}-2s_{34}s_{45}-s_{14}s_{26}
+ s_{35}\ =\ 0 \cr
&2(2s_{34}s_{56}s_{45}-s_{34}s_{46}-s_{56}s_{35})\cr
+&s_{14}s_{25}-s_{12}s_{45}-s_{24}s_{15}+ s_{36}\ =\ 0 \cr}
$$
and their images under cyclical permutations of the indices $1\cdots 6$.

The symplectic volume form on the algebraic manifold defined by
$I(s)=0$ expressed as
$$
dq_{a} \wedge dQ_{a}\eqno(4.13)
$$
can be simplified by noting that the $t_{ij}$ can be identified
with the differentials $ds_{ij}$ as follows; from
$$
dI(s)=\o {\pa I(s)}{\pa s_{ij}} ~ds_{ij} =\o {\pa I(s)}{\pa s_{ij}}
{}~t_{ij}
$$
then (4.11) should be compared with
$$
ds_{ij} = \o {\pa s_{ij}}{\pa q_a} dq_a
+ \o {\pa s_{ij}}{\pa Q_a} dQ_a
$$

and implies the identifications
$$
dQ_{a}=  -p_{a} \qquad or \qquad dq_{a} = P_{a}
$$
so the symplectic form (4.13) can be written simply as
$$
P_{a} \wedge dQ_a = p_{a} \wedge dq_a
$$
The commuting $q_a$ and $Q_a, a=1,2,3$ form a basis for
configuration space , whereas the commuting momentum (tangent space)
variables $p_{a}$ and $P_{a}$ are given by (4.12).

The action of the mapping class group (2.7) for $g=2$ has been computed
explicitly only on the cosines of the variables $q_a$ and $Q_a$,
on the sines it is very complicated and not useful. A simple example
however is given by the transformation $D_{2j-1,2j}, j=1,2,3$,
generated classically on $\pi_{1}(\Sg)$ by
$$t_2 \to t_1~t_2 \qquad t_6 \to t_6~{t_1}^{-1}
$$
and on the traces of holonomies by the map (canonical transformation)
which leaves invariant the volume form (4.13)
$$
D_{2j-1,2j}; q_j \to q_j + Q_j
$$
with inverse
$$
{D^{-1}}_{2j-1,2j}; q_j \to q_j - Q_j
$$
and can be identified with a Dehn twist [14] provided the $Q_a$ play the
role of the length variables and the $q_a$ are the angle variables for a
given closed path on $\Sg$.
The full action of this group is under study and will be reported
elsewhere.
\beginsection References.

\item[1] J.E.Nelson, T.Regge, Phys.Lett. {\bf B272},(1991)213.
\item[2] J.E.Nelson, T.Regge, F.Zertuche, Nucl.Phys.
{\bf B339},(1990)516: F. Zertuche, Ph.D.Thesis, SISSA (1990),unpublished.
\item[3] J.E.Nelson, T.Regge, C.M.P.{\bf 155},(1993)561.
\item[4] J.E. Nelson and T.Regge, Phys.Rev. {\bf D50},(1994)5125.
\item[5] J.E.Nelson, T.Regge, Nucl.Phys. {\bf B328},(1989)190.
\item[6] L.F.Urrutia and F.Zertuche, Class.Qu.Grav.{\bf 9},(1992)641.
\item[7] S.Carlip and J.E.Nelson, Phys.Lett. {\bf B324},(1994)299;
gr-qc/94110312, to appear Phys.Rev. {\bf D}
\item[8] J.E.Nelson, T.Regge, C.M.P.{\bf 141},(1991)211.
\item[9] J.S.Birman, Comm.Pure.Appl.Math.{\bf 22}(1969)213.
\item[10] K.Ezawa, Osaka preprint OU-HET-183 (1993),hep-th/9311103
\item[11] J.Louko and D.Marolf, Class.Qu.Grav. {\bf 11} (1994)311.
\item[12] R.Loll, "Independent Loop invariants for 2+1 Gravity",
gr-qc/9408007.
\item[13] A.Ashtekar and R.Loll, Class.Qu.Grav. {\bf 11} (1994)2417.
\item[14] M.Rasetti, in "Symmetries in Science III" (eds. B.Gruber and
F.Iachello) Plenum 1989.

\vfill\eject

\bye